\documentclass[runningheads]{llncs}
\usepackage{amsmath}
\usepackage[T1]{fontenc}
\usepackage{graphicx}
\usepackage{amssymb}
\usepackage{comment}
\usepackage{mathtools}
\usepackage{multirow}
\usepackage{booktabs}
\usepackage{hyperref}
\footnotesize
\tiny
\begin{document}
\title{A Unified Framework for Foreground and Anonymization Area Segmentation in CT and MRI Data}
\titlerunning{Foreground \& Anonymization Segmentation Framework}
%
\author{Michal Nohel \inst{1,2,3},
    Constantin Ulrich \inst{1,5,6},
    Jonathan Suprijadi \inst{1}, 
    Tassilo Wald \inst{1,7},
    Klaus Maier-Hein \inst{1,4,6,7} 
}
\authorrunning{Nohel et al.}

%
\institute{ 
German Cancer Research Center (DKFZ), Heidelberg, Division of Medical Image Computing, Germany\and
Department of Biomedical Engineering, Faculty of Electrical Engineering and Communication, Brno University of Technology, Brno, Czech Republic\and
University hospital Ostrava, Department of Deputy director for science, research and education\and
Pattern Analysis and Learning Group, Department of Radiation Oncology, Heidelberg University Hospital, Heidelberg, Germany\and
National Center for Tumor Diseases (NCT), NCT Heidelberg, A partnership between DKFZ and University Medical Center Heidelberg \and
Medical Faculty Heidelberg, University of Heidelberg, Heidelberg, Germany \and
Faculty of Mathematics and Computer Science, Heidelberg University, Germany 
\email{constantin.ulrich@dkfz-heidelberg.de}
}
\maketitle              
\begin{abstract}
\sloppy 
This study presents an open-source toolkit to address critical challenges in preprocessing data for self-supervised learning (SSL) for 3D medical imaging, focusing on data privacy and computational efficiency. The toolkit comprises two main components: a segmentation network that delineates foreground regions to optimize data sampling and thus reduce training time, and a segmentation network that identifies anonymized regions, preventing erroneous supervision in reconstruction-based SSL methods. Experimental results demonstrate high robustness, with mean Dice scores exceeding 98.5 across all anonymization methods and surpassing 99.5 for foreground segmentation tasks, highlighting the toolkit's efficacy in supporting SSL applications in 3D medical imaging for both CT and MRI images. The weights and code is available at \href{https://github.com/MIC-DKFZ/Foreground-and-Anonymization-Area-Segmentation} {https://github.com/MIC-DKFZ/Foreground-and-Anonymization-Area-Segmentation}.
\end{abstract}

\section{Introduction}
Deep learning has achieved remarkable success in medical image analysis. Groundbreaking work in segmentation has greatly enhanced treatment planning and monitoring outcomes \cite{3671-009,3671-013,3671-004}, while advances in detection models have improved diagnostic \cite{3671-014}. Despite these advancements, progress in deep learning for medical imaging remains constrained by the scarcity of annotated data, which is essential for training robust models. Self-supervised learning (SSL) offers a solution to this challenge by leveraging vast amounts of unannotated medical images, thus potentially improving all downstream tasks. However, there are two critical limitations hindering the effective use of unannotated medical data within the community.

Firstly, a lot of volume in 3D medical images contains air, which increases training inefficiencies when sampled compared to traditional 2D natural images. This makes it essential to efficiently filter and use only those parts of the images that contribute valuable information, avoiding unnecessary computational costs. 
Secondly, data is artificially altered due to privacy concerns; publicly accessible medical datasets are often anonymized, with sensitive areas like the face, ears, and even the top of the head removed or blurred. Such alterations create challenges for deep learning applications, particularly in generative models like GANs, diffusion models, or autoencoders, which rely on original images for accurate supervision. When these images are artificially distorted, networks may receive erroneous signals. For SSL methods focused on reconstruction, identifying and excluding these altered areas from loss calculation becomes especially critical.
To improve 3D medical SSL applications, we introduce an open-source toolkit designed to preprocess data for SSL tasks. The toolkit includes two key components: (1) a network that segments all foreground regions within a CT or MR scan, allowing models to selectively sample patches from anatomical regions, thereby reducing training time; and (2) a network that identifies anonymized regions in CT and MRI images, enabling the exclusion of these regions from loss calculations to avoid incorrect supervision signals. 

\begin{figure}[b]
    \centering 
    \includegraphics[width=\linewidth]{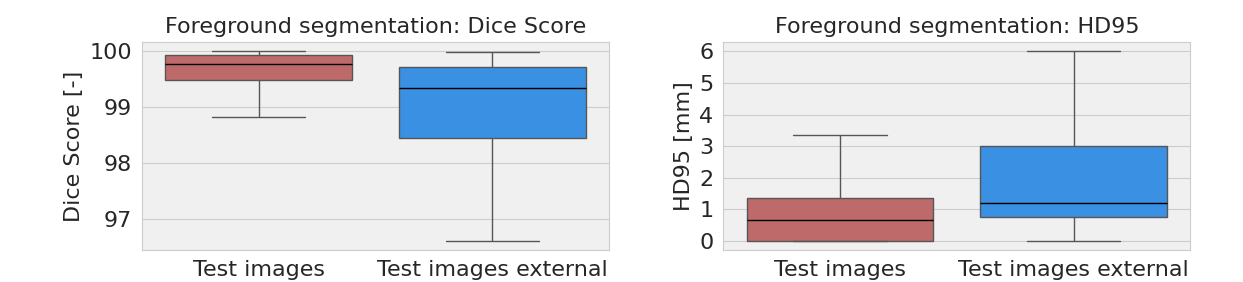}
    
    \caption{Boxplot results illustrating the Dice Similarity Coefficient and HD95 for foreground segmentation, shown for both the test portion of the training dataset and the external test dataset.}
    \label{3671-fig-01_result_foreground_segm}
\end{figure}

\section{Materials and methods}
\label{3671-sec-methods}
\paragraph{Anatomical foreground segmentation} 
\label{3671-sec-anat_forg_segm}
We assembled a diverse collection of CT and MRI images to construct a robust training dataset for anatomical foreground segmentation. This dataset includes the set of publicly available datasets introduced in Tab.~\ref{3671-tab-overview_of_datasets}.

In total, the dataset comprises 3299 3D images: 1899 CT and 1400 MRI scans. The TotalSegmentator MRI dataset was exclusively reserved for testing MRI images, and the MAAL dataset for testing CT images, while the remaining datasets were evenly split 50:50 into training and testing. This resulted in 1483 training and 1816 testing images.

A multi-step approach was applied to obtain anatomical foreground masks for both CT and MRI images. For CT images, the TotalSegmentator tool \cite{3671-004} was utilized in ``body mode'' \ to generate initial segmentations, which were manually refined using the MITK \cite{3671-007} and ITK-SNAP \cite{3671-008} software tools to ensure accurate foreground delineation by filling gaps, removing errors, and adding missing regions. On average, correcting a single 3D CT image took only a few minutes. Since TotalSegmentator was only trained on CT images, it did not perform well on MRI images. Therefore, we annotated a smaller subset of images from each dataset using intensity thresholding and manual correction. This initial segmentation process was time-consuming and required significant effort to ensure accurate delineations. Once the initial network was trained on this subset, it was used to predict segmentations for the remaining MRI images, requiring only minor manual refinements, significantly reducing the time to just a few minutes per 3D image. Finally, we trained a nnU-Net on all training data including both modalities, increasing the patch size to $192\times192\times192$ to reduce inference time and applying z-score normalization to handle CT and MRI images with a single network.

\begin{table}[t]
	\caption{Overview of all datasets used for foreground prediction. * A small number of images from the original set of 1,228 CT and 298 MRI were excluded due to file corruption or incompatibility with MITK and ITK-SNAP.}
	\label{3671-tab-overview_of_datasets}
        \resizebox{1\textwidth}{!}{
	\begin{tabular*}{\textwidth}{l@{\extracolsep\fill}lr}
		\hline
            Dataset & \# Images & Modalities \\
            \hline
        LACS \cite{3671-001} & 30 & cardiac MRI \\
		  ACDC \cite{3671-002}  & 300 & cine MRI \\
		AMOS \cite{3671-003} & 600 & 500 CT \& 100 MRI (no further information given) \\
        TotSeg \cite{3671-004} & 1360* & 1092 CT \& 268 MRI (no further information given)\\
        HanSeg \cite{3671-005} & 84 & 42 CT \& 42 T1w MRI \\
        OASIS3 \cite{3671-006} & 860 & 200 CT, 200 T1w,  200 T2w, 100 Angio \& 160 FLAIR\\	
        MAAL \cite{3671-011} & 65 & CT\\
		\hline
	\end{tabular*}}
\end{table}

\begin{figure}[b]
    \centering 
    \includegraphics[width=\linewidth]{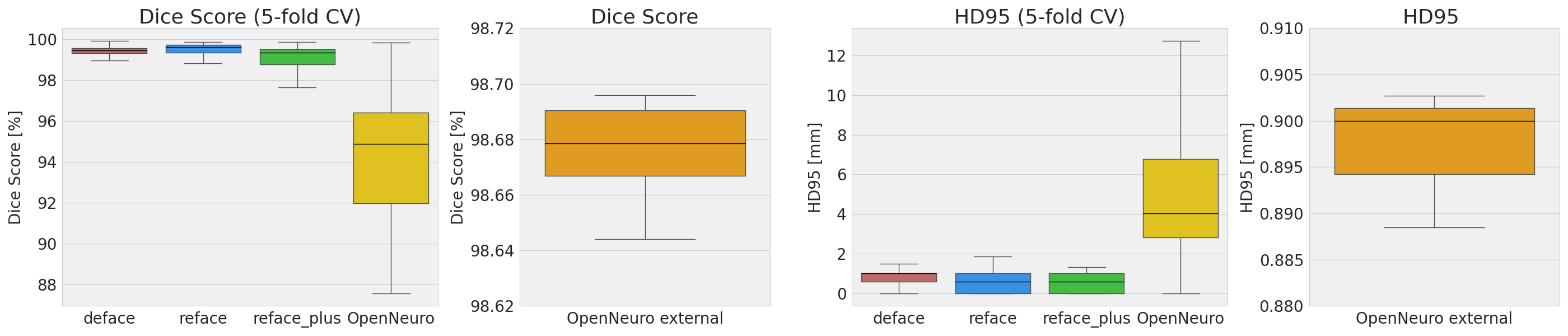}
    
    \caption{Boxplot examples illustrating the Dice coefficient and HD95 for segmentation of deface area across different anonymization modes and the external OpenNeuro dataset.}
    \label{3671-fig-02_result_deface_segm}
	
\end{figure}

\paragraph{Deface area segmentation} 
\label{3671-sec-def_area_segm}
The publicly available OASIS 3 dataset \cite{3671-006} was used as a training dataset for the segmentation of the anonymized facial parts. We selected a total of 1,005 scans of this database, comprising 319 T1-weighted (T1w) images, 329 T2-weighted (T2w) images, 68 angiographic images, 130 FLAIR images, and 159 CT images. These images, originally non-anonymized, were subjected to artificial anonymization using the AFNI (Analysis of Functional NeuroImages) software \cite{3671-010}. \par

AFNI provides three anonymization methods: deface, reface, and reface plus. Deface sets voxel values in the face and ears to zero, reface replaces them with a blurred generic face and blurred ears, and reface plus also applies blurring to the outer skull.
Each image was processed with all three methods, resulting in a comprehensive dataset of 3015 anonymized images with corresponding masks and different anonymization schemes. 

The dataset was further extended with data from the OpenNeuro dataset collection \cite{3671-012}, which is particularly valuable due to its diverse MRI head scans in different acquisition sequences and accessible licensing, making it suitable for SSL training. This collection employed different anonymization methods, with some datasets providing original deface masks originating from the  \emph{Defaced} method \cite{3671-016,3671-017,3671-018,3671-019}. There were a total of 107 MRI images in this OpenNeuro dataset, including 78 T1w and 29 T2w images. We then trained a nnU-Net on this dataset and we measured both the Dice score and 95\% Hausdorff Distance (HD95) on a five-fold cross-validation. 

In one dataset within the OpenNeuro collection, 14 images were anonymized through blurring; however, no ground truth masks were provided \cite{3671-015}. To assess the network's performance, we manually annotated these images, which were anonymized following the \emph{Reface Plus} method, blurring the face, skin, and ears. We used this OpenNeuro dataset as an external testing dataset.

\section{Results}
\label{3671-sec-results}
\paragraph{Anatomical foreground segmentation} 
\label{3671-sec-anat_forg_segm_results}
The trained model was evaluated on two test datasets: the test splits of the in-distribution training datasets, and the external datasets. The results are presented as boxplots in Fig.~\ref{3671-fig-01_result_foreground_segm}. On the test split of the training dataset, the model achieved a mean Dice coefficient of 99.56 and a median Dice coefficient of 99.76. The mean Dice coefficient for the external test dataset was 98.57, with a median of 99.34. Regarding HD95, the test portion of the training dataset showed a mean value of 1.38 and a median of 0.67, while the external test dataset yielded a mean HD95 of 2.89 and a median of 1.19. The model performs the task with near-perfect accuracy and demonstrates notable robustness, even on datasets outside the training data distribution.
\begin{figure}[t]
    \centering 
    \includegraphics[width=\linewidth]{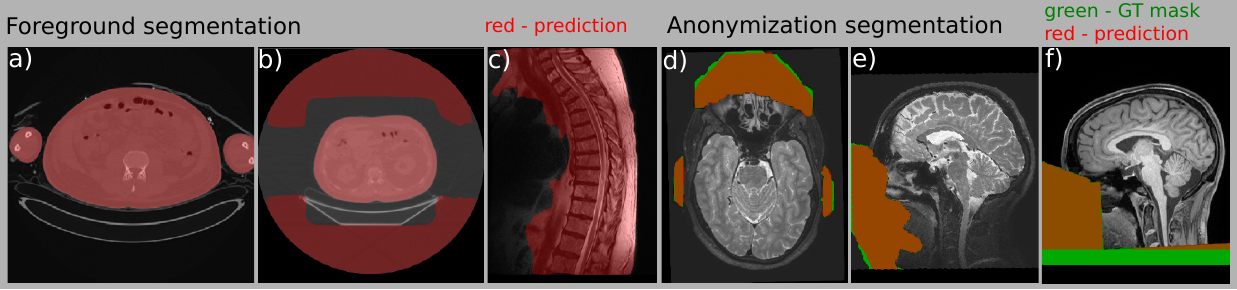}

    \caption{Example predictions for foreground (FG) segmentation and anonymization segmentation. While in the majority of cases, the FG segmentation is almost perfectly solved (a), we observed a rare failure case for volumes with an artificial boundary between actual air and a region with constant values introduced during image reconstruction(b). Further we observed problems in images without a sharp contrast between the background and surrounding body regions, where even a human would not be able to deliniate the background (c). Lastly, the ground truth masks for the Deface anonymization (green) in the Open Neuro dataset were occasionally ambiguous. While the network accurately predicted the removed portions of the head in the foreground (d-e), we observed in some cases an ambiguity of the anonymization ground truth mask in the background (f).}
	\label{3671-fig-03}
\end{figure}

\paragraph{Deface area segmentation} 
\label{3671-sec-def_area_segm_result}
We employed a 5-fold cross-validation to the model trained for deface segmentation. Results for each anonymization type, as well as for the external dataset with manually created segmentation, are presented in Fig.~\ref{3671-fig-02_result_deface_segm} and Tab.~\ref{3671-tab-results_deface_segm}. Again, the model solves this task almost perfectly and shows robust performance on the holdout test dataset.

\begin{table}[t]
	\caption{Mean, standard dev.,  and median values of the Dice Similarity Coefficient and HD95 for each anonymization type on the test dataset, including results for the external blurred face dataset.}
	\label{3671-tab-results_deface_segm}
    \resizebox{1\textwidth}{!}{
	\begin{tabular*}{\textwidth}{l@{\extracolsep\fill}llll}
            \hline
            & \multicolumn{2}{c}{Dice Coefficient [\%]} & \multicolumn{2}{c}{HD95 [mm]} \\
		\hline
            & mean (std) & median & mean (std) & median  \\
            \hline
		Deface & 99.05 $\pm$ 4.29 & 99.46 & 5.08 $\pm$ 43.44 & 1.00  \\
		  Reface & 99.50 $\pm$ 0.46& 99.63 & 0.68 $\pm$ 3.85 & 0.59 \\
		  Reface Plus & 99.01 $\pm$ 1.92 & 99.34 & 0.72 $\pm$ 3.43 & 0.59  \\
        OpenNeuro  & 92.90 $\pm$ 7.14 & 94.88 & 13.57 $\pm$ 29.16 & 4.03\\	
		OpenNeuro - external & 98.67 $\pm$ 0.05& 98.67 & 0.90 $\pm$ 0.01 & 0.90\\	  	
		\hline
	\end{tabular*}}
	
\end{table}

\section{Discussion}
\label{3671-sec-discussion}
\label{3671-sec-anat_forg_segm_discussion}
The foreground segmentation network demonstrated robust performance across all test data, achieving near-perfect results, even on external datasets. However, certain edge cases revealed limitations that merit discussion. Specifically,  we observed some rare cases as illustrated in Fig.~\ref{3671-fig-03} b and c), where the network produced an incorrect prediction. This error, visualized in Fig.~\ref{3671-fig-03} b), typically occurred in images with a large field of view lacking foreground content that exceeded the patch size, when there was also a boundary between actual air and a region with constant values—likely introduced during image reconstruction or processing. The network misinterpreted this boundary as foreground.

\label{3671-sec-def_area_segm_discussion}

The deface segmentation network demonstrated highly robust performance across all anonymization types. However, as shown in Fig.~\ref{3671-fig-02_result_deface_segm},  the segmentation accuracy for the OpenNeuro is slightly lower. This increased variability arises from ambiguities in the available masks. In defaced cases, we observed instances where the anonymization mask prediction in the foreground was accurate, but the boundaries of the ground truth mask in the background remained ambiguous, leading to mis-segmentation, as illustrated in Fig.~ \ref{3671-fig-03} f). These failure cases were extremely rare and occurred solely in the background of the volume, presenting no limitations to the toolkit's functionality or intended purpose. At worst, the network might mistakenly sample from the background or omit a small background region in the loss calculation for potential SSL applications. Overall, we have introduced a highly accurate and versatile tool for segmenting anatomical foreground and anonymized regions, providing a valuable resource that particularly benefits SSL methods in medical imaging. \par
\vspace{10pt}

\sloppy 
\noindent \textbf{Acknowledgement.} This article has been produced with the financial support of the European Union under the LERCO project number \texttt{CZ.10.03.01/00/22\_003/0000003} via the Operational Programme Just Transition.


\bibliographystyle{splncs04}
\bibliography{BibFile}
\end{document}